\documentclass[reprint,amsmath,amssymb,aps]{revtex4-1}
\usepackage{float}
\usepackage[colorlinks, linkcolor= blue, citecolor = blue, urlcolor=blue]{hyperref}
\usepackage{graphicx}
\usepackage{dcolumn}
\usepackage{bm}
\begin{document}

\preprint{APS/123-QED}

\title{Enhanced Activity Reduces the Duration of Intermittent L\'evy Walks\newline in Bacterial Turbulence}
\author{G. Dhananjay$^{\rm{1}}$, M. Hemlata$^{\rm{1}}$, Saravanan Matheshwaran$^{\rm{2}}$, and Sivasurender Chandran$^{\rm{1}}$}

\email{schandran@iitk.ac.in}
\affiliation{$^{\rm{1}}$ Department of Physics, Indian Institute of Technology, Kanpur\\
$^{\rm{2}}$ Department of Biological Sciences and Bioengineering, Indian Institute of Technology, Kanpur}


\begin{abstract}
Dense bacterial suspensions display collective motion exhibiting coherent flow structures reminiscent of turbulent flows. In contrast to inertial turbulence, understanding the microscopic dynamics of bacterial fluid elements undergoing collective motion is in its incipient stages. Here, we report experiments revealing correlations between the microscopic dynamics and the emergence of collective motion in bacterial suspensions. Our detailed analysis of the passive tracers and the velocity field of the bacterial suspensions allowed us to systematically correlate the Lagrangian and the Eulerian perspectives. Bacteria within the collective dynamics revealed initial ballistic dynamics followed by intermittent L\'evy walk before the eventual decay to random Gaussian fluctuations. Intriguingly, the persistence length and time of the fluid motion decrease with an increase in the activity, which, in turn, reduces the duration of L\'evy walk. Our results reveal transitions in microscopic dynamics underlying the bacterial turbulence and their control via the intrinsic time scales set by the effective activity of the flow.
\end{abstract}
\maketitle
Examples of collective motion could be witnessed across various scales ranging from the picturesque flocking of birds, and huddling of penguins, to the crawling of cells and the swarming of bacteria \cite{Doostmohammadi2018, RevModPhys.85.1143, Ramaswamy_2017,Yeomans2014, TONER2005170,doi:10.1073/pnas.1107583108,doi:10.1146/annurev-conmatphys-070909-104101,doi:10.1073/pnas.1005766107,PhysRevLett.110.228102}. Despite the apparent differences in their composition and length scales, dedicated efforts over the last couple of decades revealed many similarities in their statistical properties like the scale dependence of spatiotemporal correlations, and the nature of interactions \cite{Doostmohammadi2018, RevModPhys.85.1143, Ramaswamy_2017}. In general, a complex interplay of several factors including decision-making based on environmental cues, biochemical gradients, and physical interactions may manifest into macroscopic collective motion \cite{Doostmohammadi2018, RevModPhys.85.1143, Ramaswamy_2017,Yeomans2014, TONER2005170,doi:10.1073/pnas.1107583108,doi:10.1146/annurev-conmatphys-070909-104101,doi:10.1073/pnas.1005766107,PhysRevLett.110.228102,PhysRevLett.98.158102, PhysRevLett.93.098103, Gachelin_2014, PhysRevLett.98.158102,PhysRevLett.109.248109, PhysRevLett.110.228102,2022, 10.3389/fmicb.2021.715220,Aranson_2022,Cisneros}. Here, the absence of cognitive ability in lower forms of life like microbial populations and living polymers (microtubules) reduce the overall complexity of the problem and thus, are excellent model systems to explore the physics underlying the rich dynamics of collective motion \cite{PhysRevLett.98.158102, PhysRevLett.93.098103, Gachelin_2014, PhysRevLett.98.158102,PhysRevLett.109.248109, PhysRevLett.110.228102,2022, 10.3389/fmicb.2021.715220,Aranson_2022,Cisneros,PhysRevLett.109.248109,microtubule1,microtubule2}. For instance, the collective behavior observed in bacterial fluids is broadly understood on the basis of interplay between simple physical interactions, force gradients, and local alignment rules \cite{VICSEK201271, Doostmohammadi2018, RevModPhys.85.1143, Ramaswamy_2017,PhysRevLett.75.4326,TONER2005170,https://doi.org/10.48550/arxiv.1812.00310,Yeomans2014}. While the orientational order within the bacterial communities may underlie excluded volume interactions between the entities, the hydrodynamic interactions between the neighboring aligned swimmers manifest into collective dynamics including large-scale coherent flow structures like vortices, traveling waves, and jets \cite{Qi2022,PhysRevLett.99.058102}. Dynamic similarities of such coherent flow structures with that of fluid flow in inertial turbulence, especially the emergent power laws in the length scale dependence of the energy spectrum \cite{doi:10.1146/annurev-conmatphys-082321-035957,Alert2020}, inspired coinage of the word active turbulence to describe their flow dynamics \cite{doi:10.1073/pnas.1202032109,Qi2022}. Thanks to the intense efforts at the continuum scales, the patterns formed during collective motion can be simulated to a greater extent by just invoking the competition of active stresses and dissipation \cite{Dunkel_2013, PhysRevLett.110.228102}. 

The success in capturing the overall flow dynamics allows us to delve deeper into the statistical properties of individual entities involved in collective motion. Several aspects need our attention: Does the dynamics of individual entities change during the emergence of collective motion?  Experiments on swarming colonies on a surface suggest that the individual entities exhibit L\'evy walks \cite{Ariel2015,doi:10.1073/pnas.1107046108}. Recent simulations, solving Toner-Tu-Swift-Hohenberg equations, capturing the Lagrangian characteristics of the active fluids indicated the presence of initial ballistic motion followed by anomalous diffusion and L\'evy walks \cite{PhysRevLett.127.118001}. This observation was in stark contrast to the inertial turbulence as the initial ballistic dynamics is expected to become Brownian at later stages \cite{Xia2013}. This suggests that the presence of anomalous diffusion and L\'evy walks may allow distinguishing the active and inertial turbulence \cite{PhysRevLett.127.118001}. However, to date, there are no experiments validating such a transition from ballistic to anomalous diffusion. Clearly, it is essential to experimentally characterize the statistics of microscopic dynamics and interactions to better understand the interactions/rules governing collective motion.

\begin{center}
    \begin{figure*}
    \centering
    \includegraphics[width = \linewidth]{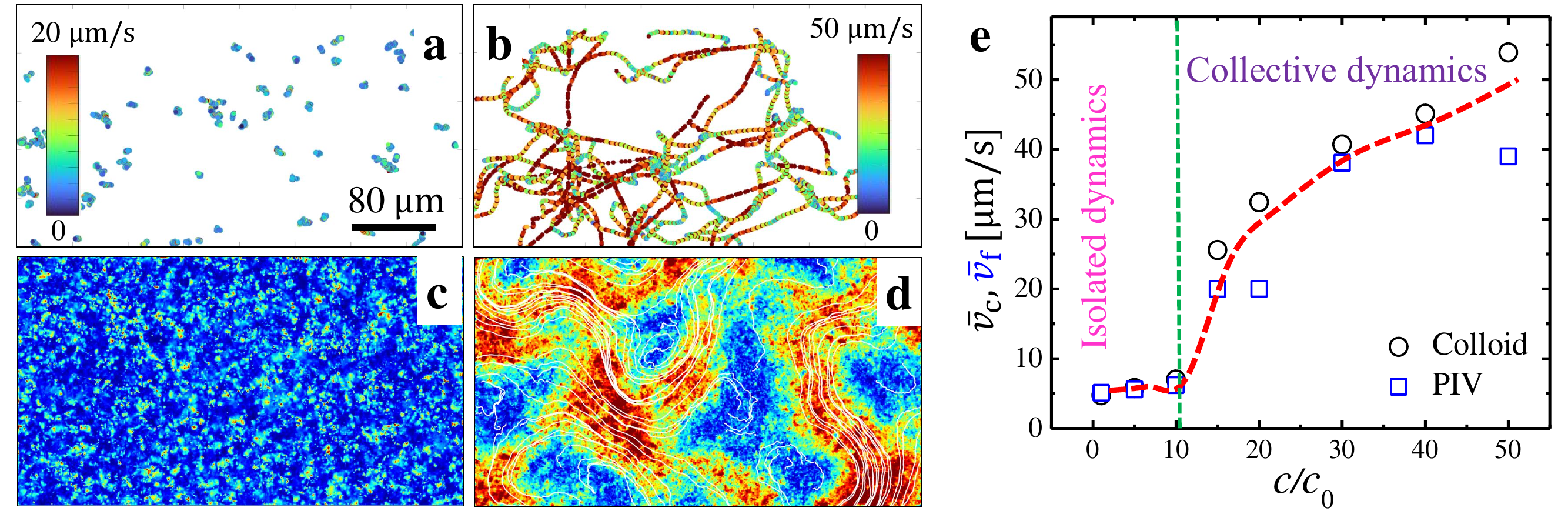}
    \caption{Equivalence of Lagrangian and Eulerian description of the flow: (a,b) shows the trajectories of colloidal particles in a bacterial bath of concentrations $c = c_{\rm{0}}$ and $c = 30c_{\rm{0}}$, respectively. Color scales represent the instantaneous velocity of the colloids. (c,d) shows the velocity fields with overlaid streamlines generated using the PIV lab. The color scale of panels (c) and (d) are similar to panels (a) and (b), respectively. (e) Mean velocities obtained from the colloids and from the PIV data are plotted as a function of bacterial  concentration. The transition after $c = 10c_{\rm{0}}$ represents the onset of collective dynamics. Open symbols and dashed line in panel (e) respectively represent experimental data and guide to the eyes.}
    \label{fig:1}
    \end{figure*}   
\end{center}

Addressing these pertinent aspects, we report experiments correlating the microscopic dynamics and the collective dynamics observed in dense bacterial suspensions. We systematically controlled the concentration of bacteria in a sessile droplet allowing us to pan the concentration-induced transition from isolated to collective dynamics. Our experiments probe the two-dimensional collective dynamics by focusing only on the air-water interface. Investigating the aforementioned fluids through the utilization of tracer particles will provide valuable insights into the intricate mechanisms governing these flows \cite{PhysRevE.105.054605, Angelani_2019, Dolai, Dikshit, PhysRevLett.103.198103, PhysRevLett.116.068303,2022,PhysRevLett.84.3017,C1SM05260H,PhysRevLett.116.068303}. We used passive colloids to track the underlying fluid flow and the resultant Lagrangian statistics are compared with the overall two-dimensional flow of the bacterial fluid. The one-to-one correspondence of the Lagrangian statistics of the tracers and the Eulerian flow field in our experiments suggest that the colloidal particles act purely as tracers. Thus, modeling the geometry of colloidal trajectories allowed us to investigate the microscopic nature of the collective dynamics. Our experiments demonstrate, for the first time, that at the microscopic level, the individual entities of the bacterial fluid display a short time ballistic dynamics followed by intermittent L\'evy walks and a Brownian motion at later stages. Intriguingly, we observe a reduction in the duration of L\'evy walks with an enhancement in the overall activity of the flow. Our results highlight the need for a better understanding of the collective dynamics in bacterial systems.  

We use suspensions of \emph{Bacillus~subtilis~IITKSM1} \cite{Murugan2019-gy, Murugan2022.02.15.480532}. The cell body is approximately 1 $\mu$m in diameter and 6 $\mu$m in length. Polystyrene (PS) colloids, procured from Sigma Aldrich, of diameter 3 $\mu$m are used as tracer particles. Bacterial suspension with concentrations $c_{\rm{0}}\approx 10^8$ cells/mL is used as the base solution. All other concentrations ($c$) are obtained by centrifuging the base solution and redispersing the settled part in appropriate amounts of nutrient solution. For our systematic study, we vary $c$ from 1$c_{\rm{0}}$ to 50$c_{\rm{0}}$. A drop of freshly prepared solutions (mixed with 0.04 wt.$\%$ PS colloids) placed on freshly hydrophilized coverslips (drop diameter $\approx$ 1 cm, after spreading) is probed using time-lapse optical microscopy. The dynamics of colloids are tracked using ImageJ package \cite{Schneider2012} and the underlying velocity distribution of the fluids using the MATLAB-based particle imaging velocimetry tool (PIV lab) \cite{PIV,Thielicke-2021}. Please refer to sections S1 and S2 in the supporting information (SI) for experimental details and analysis protocols, respectively.

Figure $\ref{fig:1}$ shows the colloidal trajectories spanning a duration of 10 s (a, b) and the Eulerian velocity fields (c, d) of the underlying fluid for two different concentrations: $c$ = $c_{\rm{0}}$ and $c$ = $30c_{\rm{0}}$. At lower concentrations, the dynamics of colloids (Fig.$\ref{fig:1}$(a)) are limited to length scales corresponding to their own diameter. The two-dimensional velocity distribution (Fig.$\ref{fig:1}$(c)) captures specific regions where the velocities are high indicating the movement of bacteria in those locations. On the other hand, colloids dispersed in dense bacterial suspensions show longer trajectories with approximately ten times higher instantaneous velocities (Fig.$\ref{fig:1}$(b)). The corresponding spatial distribution of velocities displays larger regions of high activity (Fig.$\ref{fig:1}$(d)), revealing collective motion. The characteristics like nonlinear velocity fields, the presence of vortices, and much longer trajectories (better mixing) suggest reminiscence to (inertial) turbulent flows. Interestingly, the streamlines (white lines in Fig.$\ref{fig:1}$(d)) reveal two important aspects: a) fluid elements span the entire system, and b) the neighboring streamlines are parallel over long distances revealing the presence of highly correlated flows. These observations suggest a concentration-dependent transition from isolated to collective dynamics \cite{PhysRevLett.98.158102, PhysRevLett.110.228102}. Quantifying these observations, in Fig.$\ref{fig:1}$(e), we show the mean velocity of the colloids ($\bar{v}_{\rm{c}}$) and of the fluid elements ($\bar{v}_{\rm{f}}$) as a function of bacterial concentration [refer Fig.S1(a)+(b) for the distribution of $\bar{v}_{\rm{c}}$ and $\bar{v}_{\rm{f}}$]. Clearly, both $\bar{v}_{\rm{c}}$ and $\bar{v}_{\rm{f}}$ increase rapidly beyond a threshold bacterial concentration ($c$ = $10c_{\rm{0}}$) capturing the emergence of collective dynamics. In addition, the fact that $\bar{v}_{\rm{c}}\approx \bar{v}_{\rm{f}}$ demonstrates that colloids simply trace the flow dynamics of the underlying fluid, hence, allowing to capture the Lagrangian statistics of the bacterial turbulence. Thus, probing the colloidal dynamics and the dynamics of the bacterial fluid allow us to obtain both the Lagrangian and the Eulerian perspectives of bacterial turbulence. 
\begin{center}
    \begin{figure}
    \centering
    \includegraphics[width = \linewidth]{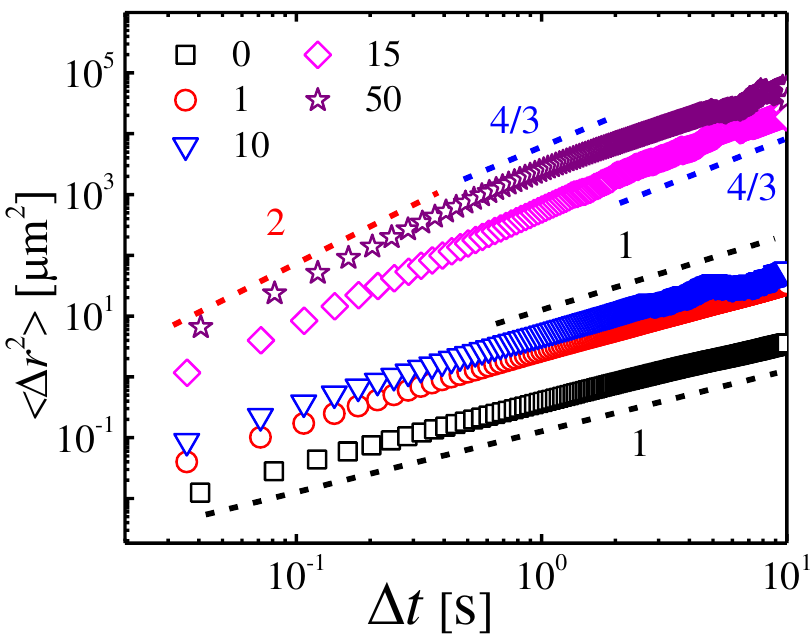}
    \caption{Capturing anomalous dynamics underlying collective motion: Double logarithmic representation of ensemble-averaged mean square displacements (MSD, $\left<\Delta r^2\right>$) of colloids as a function of time-lag ($\Delta t$). Different symbols represent different concentrations (in the units of $c/c_{\rm{0}}$) of bacterial suspensions. Exponents corresponding to different lag times and different bacterial concentrations are highlighted with dashed lines. The vertical jump after $c = 10 c_{\rm{0}}$ indicates the emergence of collective dynamics. In the collective dynamics regime, the slope of the MSD curve shows a transition from ballistic to super-diffusive nature before decaying into diffusion. With an increase in the concentration, this transition point apparently shifts toward the shorter time lags.}
    \label{fig:2}
    \end{figure}
\end{center}
\begin{figure*}
    \centering
    \includegraphics[width = \linewidth]{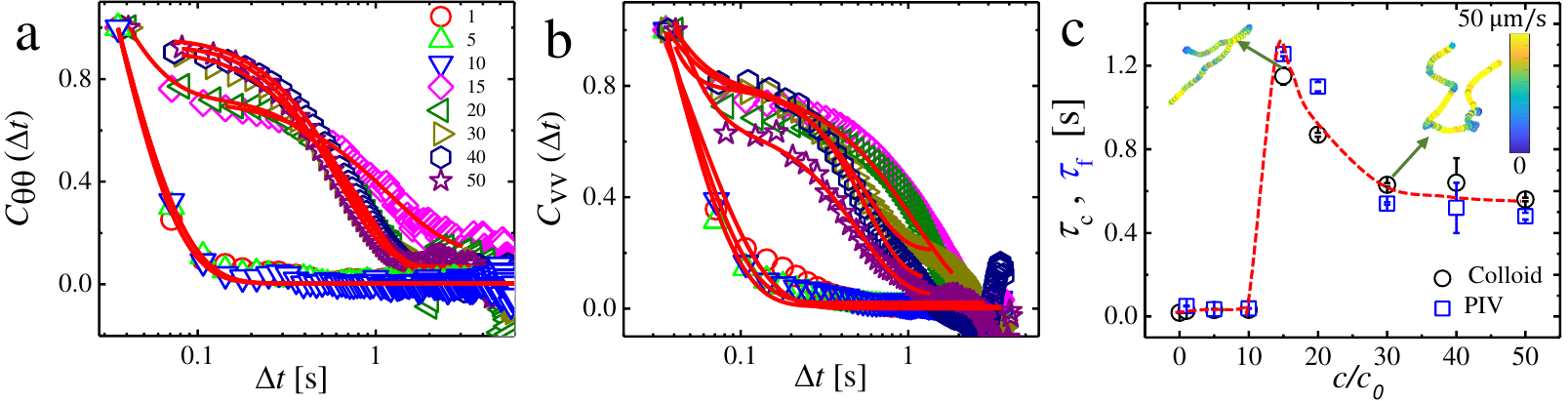}
    \caption{Flow activity sets the correlation time scales: Semi-logarithmic representation of time ($\Delta t$) dependent (a) orientational correlation function $C_{\theta \theta}$ of colloidal trajectories and (b) velocity correlation function of flow $C_{vv}$. Different symbols in (a) and (b) represent different concentrations of bacterial suspensions in the units of $c/c_{\rm{0}}$ as defined in (a). (c) Persistent time of colloids ($\tau_{\rm{c}}$) and flow correlation time ($\tau_{\rm{c}}$) as a function of bacterial concentration. Clearly, $\tau_{\rm{c}}\approx\tau_{\rm{f}}$. Also shown are the representative trajectories corresponding to $c = 15c_{\rm{0}}$ and $c = 30c_{\rm{0}}$. The persistent time is highest at the onset of collective dynamics and subsequently saturates to a lower value for higher concentrations (and correspondingly higher activities). Open symbols in all panels are experimental data. Continuous lines in (a) and (b) are best fits to the data. The dashed line in panel (c) is a guide to the eyes.}
    \label{fig:3}
\end{figure*}
To shed light on the microscopic dynamics underlying the collective motion, we quantify the trajectories of the colloidal particles using the ensemble-averaged mean squared displacement (MSD). Figure \ref{fig:2}(a) shows the double logarithmic representation of the temporal evolution of MSD of colloids dispersed in bacterial suspensions of various concentrations. Please refer Fig. S2 for the MSDs of all the concentrations used in this study. With an increase in the bacterial concentration, we observe a vertical shift of the MSDs supporting the increase in the overall velocity (see Fig.\ref{fig:1}). With increase in $c/c_{\rm{0}}$, we witness systematic variations in the exponent $\alpha$ (where, $\left<\Delta r^2\right>$ $\sim \Delta t^\alpha$). As expected, pristine colloids without any added bacteria exhibit Brownian motion ($\alpha=1$). With an increase in the bacterial concentration up to $c = 10c_{\rm{0}}$, colloids exhibit short time super diffusive ($\alpha>1$) motion, which eventually becomes random ($\alpha=1$) at longer times  \cite{D0SM00309C}. The short-time super-diffusive behavior is a manifestation of the increased rate of collisions with bacteria. For much higher concentrations, $c > 10c_{\rm{0}}$, $\alpha\approx2$ at short times suggesting a ballistic motion of colloids. At such high concentrations, the collective motion of bacteria transports colloids over large distances (see Fig.\ref{fig:1}(b)) supporting the initial ballistic motion, which is in accordance with the observations in inertial turbulence \cite{Xia2013}. In addition, we observe an intermediate region displaying anomalous diffusion ($\alpha\approx4/3$). This is a distinctive characteristic of active turbulence and is in accordance with recent experiments \cite{Ariel2015} and simulations \cite{PhysRevLett.127.118001}. Intriguingly, the time scales at which we observe $\alpha\approx4/3$ shift towards lower time scales with an increase in bacterial concentration. This is a novel observation and the microscopic reasons underlying this is not yet clear. We may conceive that the initial ballistic and the intermediate anomalous dynamics of colloids suggest persistent motions over long length and time scales. Thus, probing the role of bacterial concentration on the length and time scales of the persistent motion may improve our understanding.

We rely on the orientation correlation function ($C_{\theta\theta}$) of colloids to deduce persistence length and time [see section S3 for details]. Figure \ref{fig:3}(a) depicts the time-dependent decay of the orientation correlation functions. Building on the equivalence of the Lagrangian and the Eulerian perspectives (see Fig. \ref{fig:1}), we may expect the persistent motion of colloids to be a manifestation of the correlations in the underlying fluid. To capture this, we compute the velocity-velocity correlation function ($C_{\rm{vv}}$) and its temporal evolution. As depicted in Fig. \ref{fig:3}(b), $C_{\rm{vv}}$ systematically decayed allowing us to capture a timescale for the velocity-velocity correlation. While $C_{\theta\theta}$ and $C_{\rm{vv}}$ decay exponentially with a single relaxation time at low bacterial concentrations, colloids dispersed in dense bacterial fluids display two characteristic decays. Intriguingly, the long time decay of both $C_{\theta\theta}$ and $C_{\rm{vv}}$ show a compressed exponential with exponents ranging from 1.2 to 1.8 [see Table S1] for the tabulation of the compressed exponents]. We tentatively interpret this compressed decay of the correlation functions as a manifestation of the highly driven dynamics of the colloids and fluid elements once the collective motion sets in the system. We may conceive that the faster decay of the correlation functions is related to the fluctuation of colloids and the longer relaxation times may depend on the timescale over which flow structures stay correlated. Thus, for understanding the microscopic dynamics of the fluids, we rely only on the longer relaxation times. As depicted in Fig. \ref{fig:3}(c), the flow decorrelation time ($\tau_{\rm{f}}$) and the persistent time of the colloids ($\tau_{\rm{c}}$) are very similar, reassuring the fact that colloids simply act as tracers of the fluid flow. Intriguingly, both $\tau_{\rm{c}}$ and $\tau_{\rm{f}}$ had a maximum at the bacterial concentration $c = 15c_{\rm{0}}$, which marks the lowest concentration at which we observe the collective dynamics. Subsequently, $\tau_{\rm{c}}$ and $\tau_{\rm{f}}$ decreased and saturated at a lower value for higher bacterial concentrations. This is highly intriguing as with an increase in bacterial concentrations the overall activity, defined by the mean flow velocity, of the system increase (see Fig. \ref{fig:1}(e)) monotonously. How do we understand a decrease in the persistent time with an increase in the overall activity? For systems with $c>15c_{\rm{0}}$, the suspension becomes denser, and the overall activity (see fig.\ref{fig:1}(e)) increases. At much denser suspensions, the vortices start to mix faster and, affect the persistent motion of colloids (see the representative trajectories in Fig. \ref{fig:3}(c)). This may explain the decrease in $\tau_{\rm{c}}$ and $\tau_{\rm{f}}$ after $15c_{\rm{0}}$. Similarly, the persistence length [refer section S3 and Fig. S3(a)+(b)] of the colloid decreases for $c>15c_{\rm{0}}$, corroborating that colloids follow loop-like trajectories at higher activity. The fluid correlation length increases discontinuously around the transition from isolated to collective dynamics [refer section S3 and Fig. S4(a)+(b)]. As reported earlier \cite{PhysRevLett.109.248109,mi13050746, PhysRevLett.98.158102}, irrespective of the flow activity, the correlation length is similar for all the concentrations exhibiting collective bacterial dynamics. On the one hand, collective motion increase the persistence time (and persistence length) over orders of magnitude. On the other, we observe a novel decrease in persistence with the activity of the system. Would these changes affect the way bacteria explore the available phase space?\\
\begin{figure*}
    \centering
    \includegraphics[width = \linewidth]{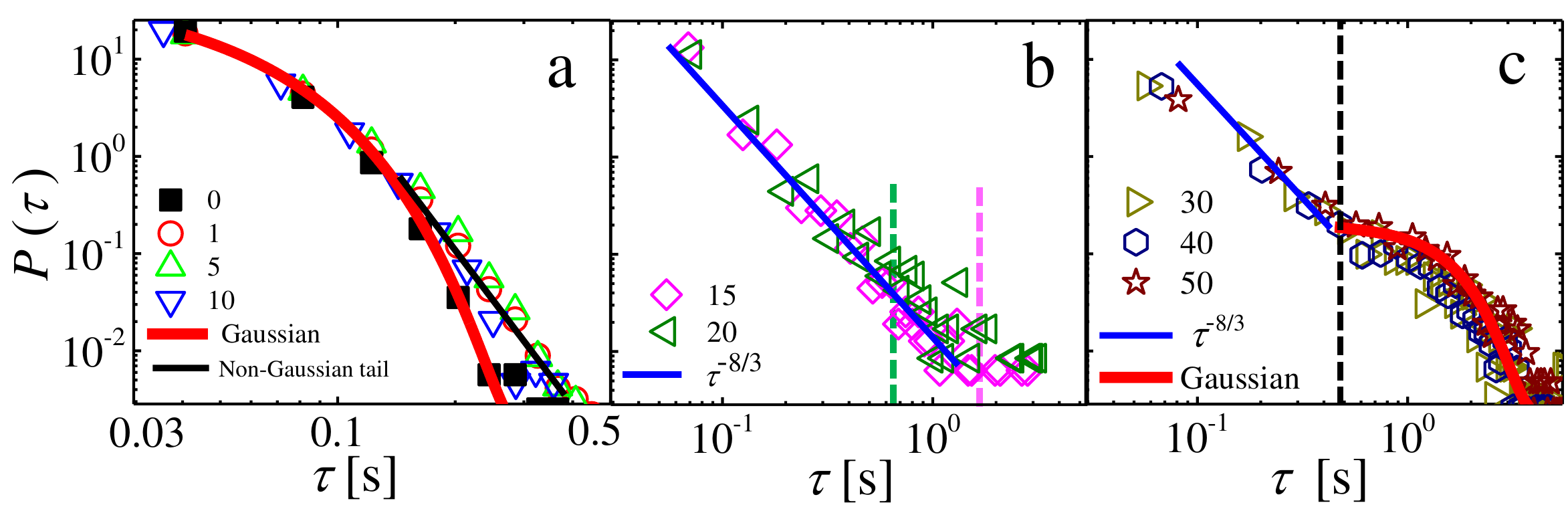}
    \caption{Transitions in the microscopic dynamics underlying bacterial turbulence: Double logarithmic representation of waiting time ($\tau$) distribution of colloids for different concentrations of bacteria. (a) For pristine colloids without any bacteria, the distribution follows Gaussian statistics capturing the expected random motion. For mild bacterial concentration ($c<10c_{\rm{0}}$) the distribution shows small deviations from Gaussian statistics at larger timescales. (b) For concentrations exhibiting collective dynamics($c= 15c_{\rm{0}}, 20c_{\rm{0}}$), the distribution shows a power law-like behavior with $\tau^{-8/3}$ scaling. This confirms the  L\'evy walk dynamics of the active bacterial suspensions. (c) For higher concentrations ( $c>20c_{\rm{0}}$), the distribution shows a Gaussian tail capturing the eventual decay to random motion. All symbols are defined in the respective panels. Vertical dashed lines in (b) and (c) reflect the respective fluid correlation times. Open symbols are experimental data and continuous lines are best fits to the data.}
    \label{fig:4}
\end{figure*} 

The long persistence time scales and length scales may suggest the presence of occasional lengthy steps, i.e., the L\'evy walks \cite{RevModPhys.87.483}. It has been observed that active suspensions exhibit a notable presence of heavy tails in their space-time distribution \cite{PhysRevLett.127.118001, Ariel2015,doi:10.1073/pnas.1107046108}, indicative of the occurrence of L\'evy-walk behavior and anomalous diffusion patterns \cite{RevModPhys.87.483}. Such lengthy steps in our experiments originate from the collective dynamics of underlying fluid and thus may reflect the spatiotemporal correlation scales of the bacterial fluid. Motivated by these results, we probe the geometrical nature of the colloidal trajectories exploring the existence of such long steps or correspondingly long times. We obtain the waiting times by dividing the trajectories into several segments based on the effective turning angle between the segments. This approach provides scope for precise analysis of the data and facilitates the calculation of waiting times corresponding to different step lengths. An angle change is recognized as a turn if $ \theta > \theta_c$, where $\theta_c = 40^\circ$ for Fig. \ref{fig:4}. Refer Fig. S5 (a)+(b) for the probability distribution of turn angles and of the distribution of waiting times corresponding to a few different turning angles. The time between two successive turns is considered as $\tau$. The probability distributions of $\tau$ for different bacterial concentrations are shown in Fig. \ref{fig:4} (a)-(c). We observe important deviations in the distribution with increasing bacterial concentration. Without any added bacteria, the colloidal tracers exhibit a Gaussian distribution of waiting times as expected for random motion. However, with added bacteria, the distribution deviates from Gaussian and has an exponential tail (see continuous lines in Fig.\ref{fig:4}(a)) at long times. This observation suggests that the diffusion of colloids at low concentrations of bacteria exhibit Brownian (see long-time behavior of MSD in Fig. \ref{fig:2}) yet non-Gaussian behavior. This is observed for tracer motion in various soft matter systems including colloids, polymers, and active suspensions \cite{PhysRevX.7.021002}. Interestingly, once the large-scale collective motion emerges (for $c>15c_{\rm{0}}$), the probability follows a power law dependence with waiting time: $P(\tau)\approx\tau^{-\gamma-1}$. The local slope analysis shows $\gamma = 5/3$ for over a decade. According to the theoretical prediction, we expect $\left<\Delta r^2\right>\sim\Delta t^{\alpha = 3-\gamma}$ for $1<\gamma<2$ \cite{RevModPhys.87.483}. With $\gamma = 5/3 $, we obtain $\alpha = 4/3 $, which is identical to the intermediate exponent in MSD [refer Fig.\ref{fig:2}]. Step length distribution follows identical behavior and thus the joint probability of waiting time and step length [refer Fig. S6] follow a linear dependence suggesting a finite L\'evy velocity. All these observations provide compelling evidence in favor of L\'evy walk. Interestingly, the most active suspensions ($c > 20c_{\rm{0}}$) exhibit a deviation from power law-like behavior at larger time scales, where a Gaussian distribution starts to set in. This highlights a transition from L\'evy walk-like dynamics to diffusion-like dynamics at longer times.

What sets these transitions from L\'evy to Gaussian for these systems, and why they are not visible in less active suspensions? We have seen that with an increase in activity, for the suspensions displaying collective motion, the flow correlation time and colloid persistence time decrease. Interestingly, the transition point for the distribution curves matches the flow correlation timescales and the persistence time of colloids (shown with the vertical dotted lines). Thus, in active suspensions, the timescale related to its flow sets an upper bound for the L\'evy walk. In less active suspensions ($c=$ 15 and 20$c_{\rm{0}}$) the flow remains correlated for a longer time than our experimental window, and thus we could not capture the transition to Gaussian statistics. With higher activity, this critical timescale appeared within our experimental window, thus, we could capture the transition. These observations clearly demonstrate that enhanced activity reduces the duration of L\'evy walk in bacterial turbulence. Earlier simulations \cite{PhysRevLett.127.118001} did not capture this transition and suggest the need for developing newer concepts and ideas. 

To conclude, we experimentally capture three different microscopic dynamic regimes underlying bacterial turbulence: initial ballistic, intermittent L\'evy walk, and the eventual Brownian dynamics. The fact that these transitions are marked by the intrinsic correlation time scales of the suspensions provides new insights into the Lagrangian description of Bacterial turbulence. The intriguing decrease in the duration of L\'evy walk with the activity of the system needs further attention. We believe that the corroboration of the Lagrangian and the Eulerian perspectives highlights that our results are applicable not only to bacterial turbulence but also to collective dynamics observed in various systems.

\textbf{Acknowledgement}: We greatly acknowledge the fruitful discussions with Rahul Pandit from IISc, Sagar Chakraborty, and Jayanta K Bhattacharjee from IIT Kanpur. We thank the initial help of Muktesh, from BSBE IIT Kanpur, in setting up the bacterial cultures. Funding support from DST-FIST (SR/FST/PS-II/2021/170(C)), SERB (SRG/2021/001276), and STARS-MoE projects is greatly acknowledged. 

\bibliography{abridgedref}
\clearpage

\begin{center}
    \textbf{\large{Supporting Information}}
\end{center}
\setcounter{figure}{0}
\setcounter{section}{0}
\renewcommand{\figurename}{Fig.}
\renewcommand{\tablename}{Table}
\renewcommand{\thefigure}{S\arabic{figure}}
\renewcommand{\thetable}{S\arabic{table}} 
\renewcommand{\thesection}{S\arabic{section}}

\section{Experimental Details}
\label{sec:S1}
For the experiments, a 10$\mu L$ aliquot is taken from an overnight bacterial culture. This sample is then inoculated in 10 mL of Luria-Bertani (LB) media. The resulting mixture is carefully placed within a falcon tube, which is loosely sealed to allow for sufficient air exchange. Subsequently, the falcon tube is inserted into a shaking incubator operating at a temperature of $37^\circ $C at 220 rpm. Afterward, the optical density (OD$_{600}$) was quantified using a Bio spectrophotometer (Eppendorf AG 223331, Hamburg). The utilization of OD$_{600}$ as a means to assess cell density in liquid culture is a common practice. The bacterium \emph{Bacillus subtilis IITKSM1} has been found to exhibit a cell density of approximately $10^8$ cells per milliliter when measured at an OD$_{600}$ of 1. In order to initiate the experimental procedure, the bacterial samples were extracted from their growth phase, characterized by an OD$_{600}$ range of 0.7-1.2.\\

Subsequently, the samples were subjected to centrifugation at a rate of 4000 rpm for a duration of ten minutes. The desired concentration can be achieved by mixing the settled part with the appropriate amounts of nutrient solution. In this study, the solution for the experiment was prepared by combining 50 $\mu L$ of the bacterial solution with 5 $\mu L$ of colloidal solution with a concentration of 0.5 $ wt.\%$. The effective concentration of colloids in the final bacterial solutions is \textit{ca}. 0.04 $ wt.\%$. From this solution, we take 10 $\mu L$ and put a drop on the glass slide. The glass slides were subjected to a thorough cleaning process utilizing the widely accepted standard RCA cleaning procedure. The cover slips utilized in this experiment have been freshly cleaned and possess hydrophilic properties. Consequently, when a drop is placed on the cover slips, it exhibits radial spreading. The drop diameter is measured to be around 1 cm, which gives a nearly flat surface near the center of the drop and possesses the least curvature effects. The colloidal motion is captured through a high-resolution time-lapse optical microscope, OLYMPUS BX53M. 

\section{ Movie analysis}
\label{sec:S2}
In order to monitor and analyze the movement of colloids, we employed ImageJ. In the initial step, the experimental movie was converted into individual frames. Subsequently, these frames were further processed to generate binary files by applying an appropriate threshold. To distinguish the colloid from the background, we applied filters based on the size of the colloid. To model the geometrical nature of the colloidal trajectories, we relied on custom-made codes in MATLAB. For the Particle Image Velocimetry (PIV), we used the open-access Matlab PIV tool. The smallest pixel size for PIV was chosen to be 4 pixels. 

\section{Length and time correlations}
\label{sec:S3}
The orientational correlation of the colloid's motion is calculated as: 
\begin{equation}
    C_{\rm{\theta\theta}}(\Delta t) = \left<\cos{\theta(\Delta t)}\right>
\end{equation}
$ \textbf{R}(t) $ is the displacement vector connecting points $r(t)$ and $r(t-\delta t)$, and $\theta$ is the angle between vectors $\textbf{R}(t+\Delta t)$ and $\textbf{R}(t)$. Here, $\delta t$ is the smallest time step accessible in our experiments and $\Delta t$ is a variable and can be any multiples of $\delta t$.

\begin{equation}
    \cos{\theta(\Delta t)} = \frac{\textbf{R}(t+\Delta t)\cdot \textbf{R}(t)}{|\textbf{R}(t+\Delta t)|| \textbf{R}(t)|}
    \label{cosq}
\end{equation}
The orientational correlation of the colloid's motion with distance is calculated as 
\begin{equation}
    C_{\theta\theta}(\Delta r) = \left<\cos{\theta(\Delta r)}\right>
\end{equation}
Where $\theta$ is the angle between displacement vectors R, those are separated by a distance $\Delta r$. For the averaging, we first calculate all the possible $\Delta r$ and plotted a histogram, then averaged the values of $C_{\theta\theta}(\Delta r)$ for every bin.
The time correlation of fluid flow is calculated as follows:
\begin{equation}
    C_{vv}(\Delta t) = \frac{\left<v(r,t)\cdot v(r,t+\Delta t)\right>}{|v(r,t)|^2}
\end{equation}
Where $v(r,t)$ represents the velocity field of fluid at a position $r$, at time $t$. To improve the statistics, the average is performed for all similar $\Delta t$ values.\\
The velocity-velocity space correlation is calculated as follows:
\begin{equation}
    C_{vv}(\Delta r) = \frac{\left<v(r,t)\cdot v(r+\Delta r,t)\right>}{|v(r,t)|^2}
\end{equation}
The average is performed for all similar $\Delta r$ values. The fitting of those functions has been implemented in ORIGIN-LAB.\\

\section{Turning angle and waiting time calculation}
\label{sec:S4}
For the waiting time distribution, first, we calculated the turning angles for the trajectory. For this, we used equation \ref{cosq} and calculated the turning angles for every time step ($\delta t$). For this analysis, we used $\Delta t=\delta t$. Turns are identified using a simple threshold $\theta > \theta_{\rm{c}}$, where we used four different $\theta_{\rm{c}}$ viz 20$^\circ$, 30$^\circ$, 40$^\circ$, and 50$^\circ$. The waiting time is calculated as the time taken between those turns.

\begin{table*}[h]
\begin{center}
\begin{tabular}{ |c|c|c|  }
\hline
$c/c_{\rm{0}}$& $\beta_{cc}$ & $\beta_{vv}$\\
\hline
15 & 1.2 $\pm$0.08& 1.6$\pm$0.05\\
20 & 1.7$\pm$0.04 & 1.2$\pm$0.06\\
30 & 1.3$\pm$0.03& 1.8$\pm$0.05\\
40 & 1.4$\pm$ 0.2& 1.7$\pm$0.5\\
50 & 1.5$\pm$ 0.04& 1.5$\pm$0.06\\
\hline
\end{tabular}
\caption{Compressing exponents, $\beta_{vv}$ and $\beta_{cc}$, obtained from exponential fits of the correlation functions. $\beta_{vv}$ is defined using $C_{vv}(\Delta t) = Ae^{-(t/\tau_1)}+Be^{-(t/\tau_f)^\beta_{vv}}$. A similar equation is used for obtaining $\beta_{cc}$ from $C_{\theta\theta}(\Delta t)$}
\label{demo-table}
\end{center}
\end{table*}

\begin{figure*}
    \centering
    \includegraphics[width = \linewidth]{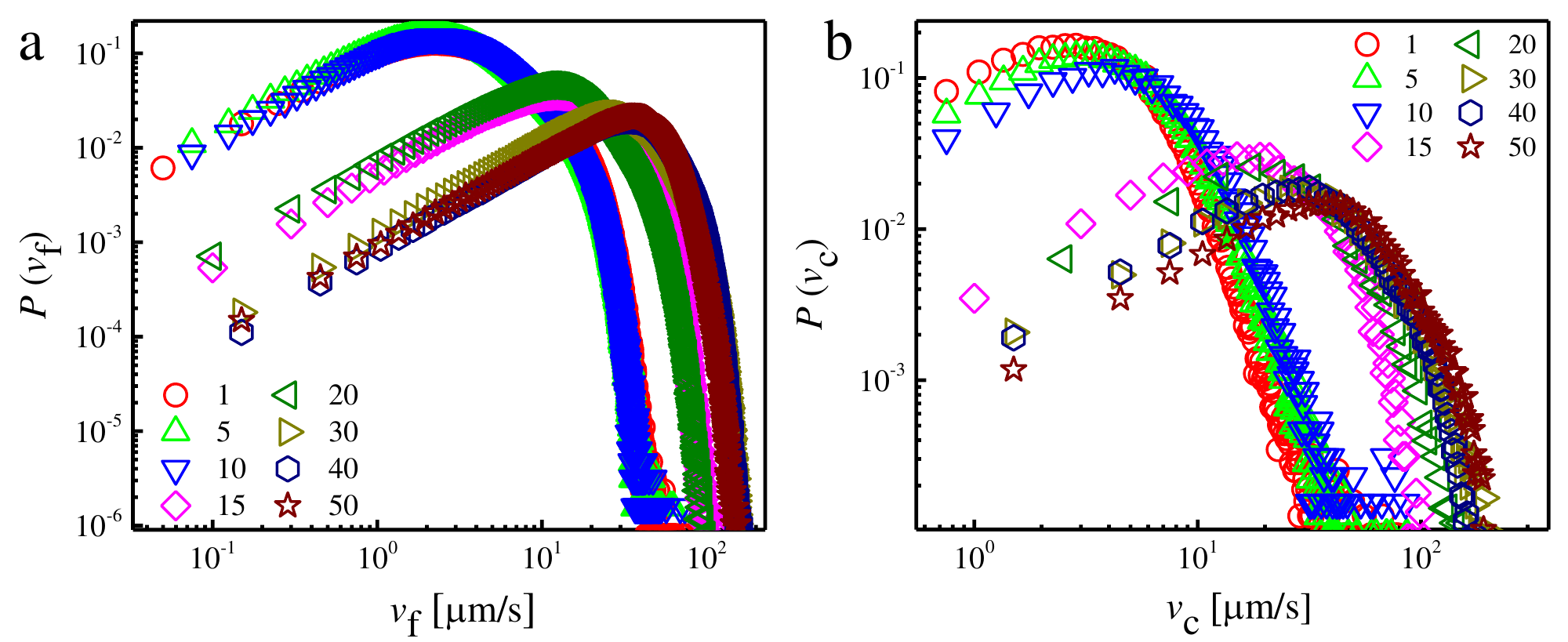}
    \caption{Probability distribution of velocities (a) from the PIV data and (b) from colloidal trajectories, for all the bacterial suspensions used in this study. Symbols define the concentration of bacterial concentrations, in the units of $c/c_{\rm{0}}$. }
    \label{fig:s1}
\end{figure*} 

\begin{figure*}
    \centering
    \includegraphics[width = 10 cm]{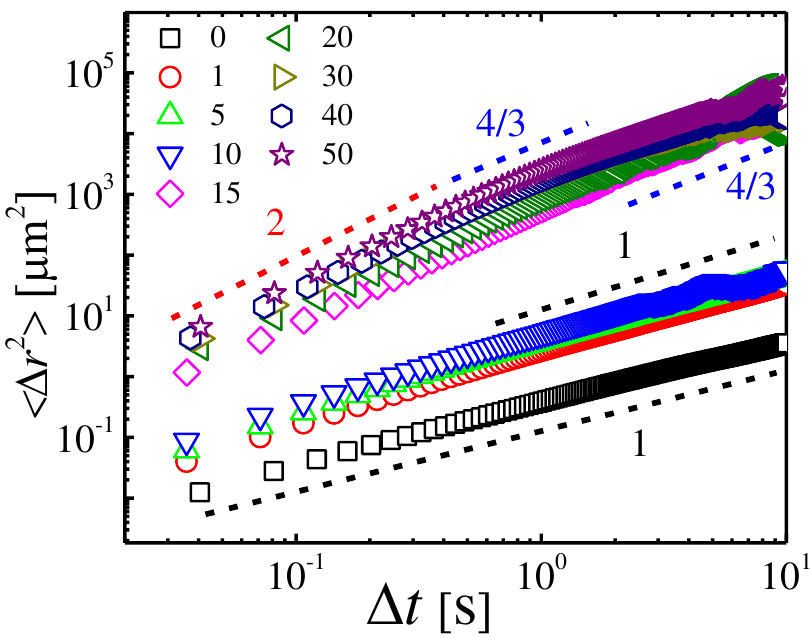}
    \caption{ Double logarithmic representations of ensemble-averaged mean square displacements (MSD, 〈$\Delta r^2$〉) of colloids
    as a function of time lag ($\Delta t$). Different symbols represent
    different concentrations of bacterial suspensions in the units
    of $c/c_{\rm{0}}$. Exponents corresponding to different lag times and different bacterial concentrations are highlighted with dashed
    lines. }
    \label{fig:s2}
\end{figure*}

\begin{figure*}
    \centering
    \includegraphics[width = \linewidth]{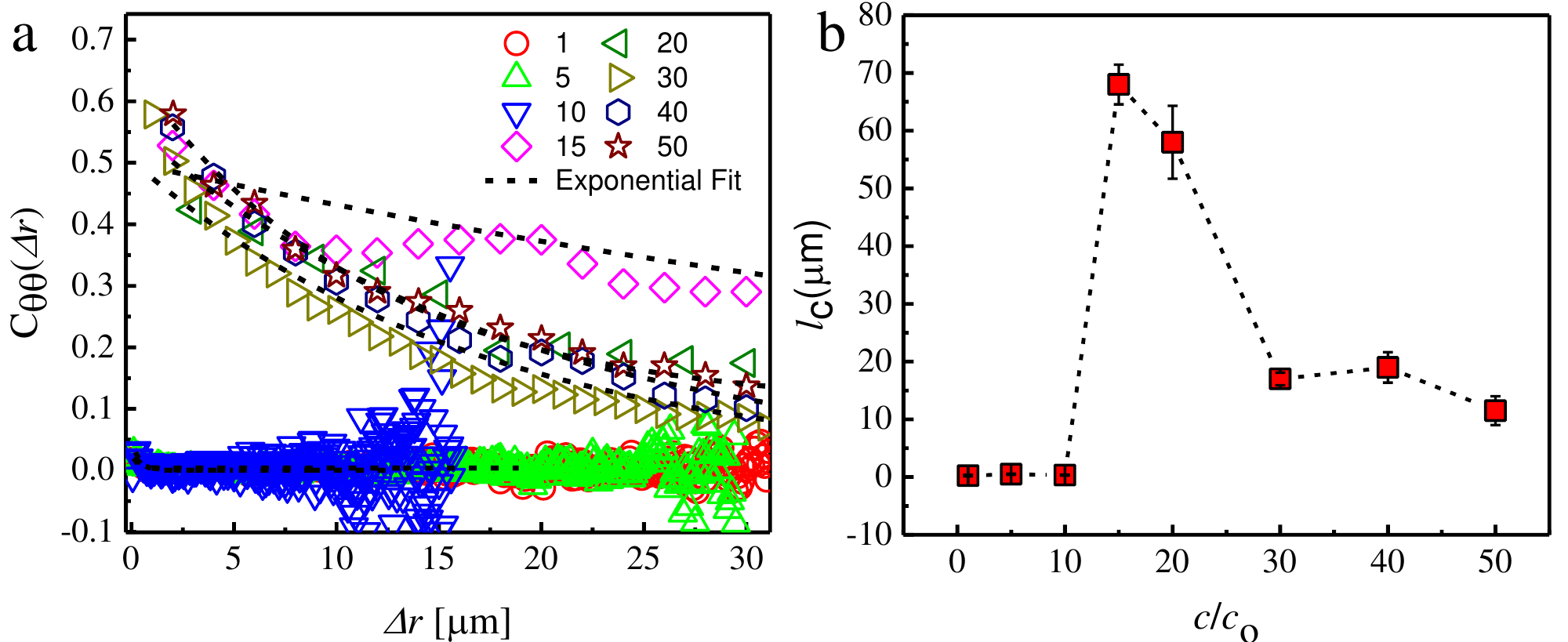}
    \caption{(a) Orientation correlation function of colloid vs $\Delta r$. (b) The persistence length of colloid ($l_{\rm{c}}$) as a function of bacterial concentration shows an increase at the $15c_{\rm{0}}$ and decreases with a further increase in the bacterial concentration. Symbols define the concentration of bacterial concentrations, in the units of $c/c_{\rm{0}}$.}
    \label{fig:s3}
\end{figure*} 

\begin{figure*}
    \centering
    \includegraphics[width = \linewidth]{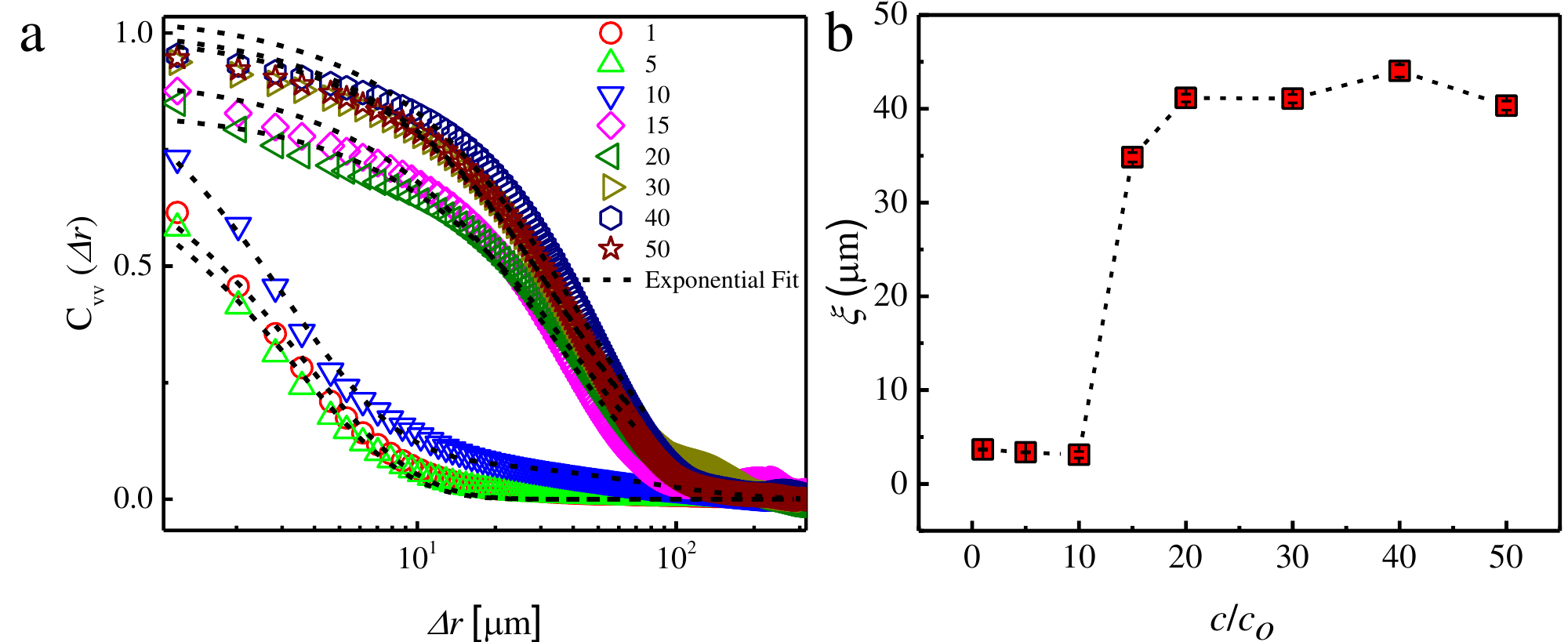}
    \caption{(a) velocity-velocity space correlation function ($C_{vv}$) vs $\Delta r$.(b) Flow correlation length scale ($\xi$) as a function of bacterial concentration, shows a plateau at higher concentrations. Symbols define the concentration of bacterial concentrations, in the units of $c/c_{\rm{0}}$. }
    \label{fig:s4}
\end{figure*}

\begin{figure*}
    \centering
    \includegraphics[width = \linewidth]{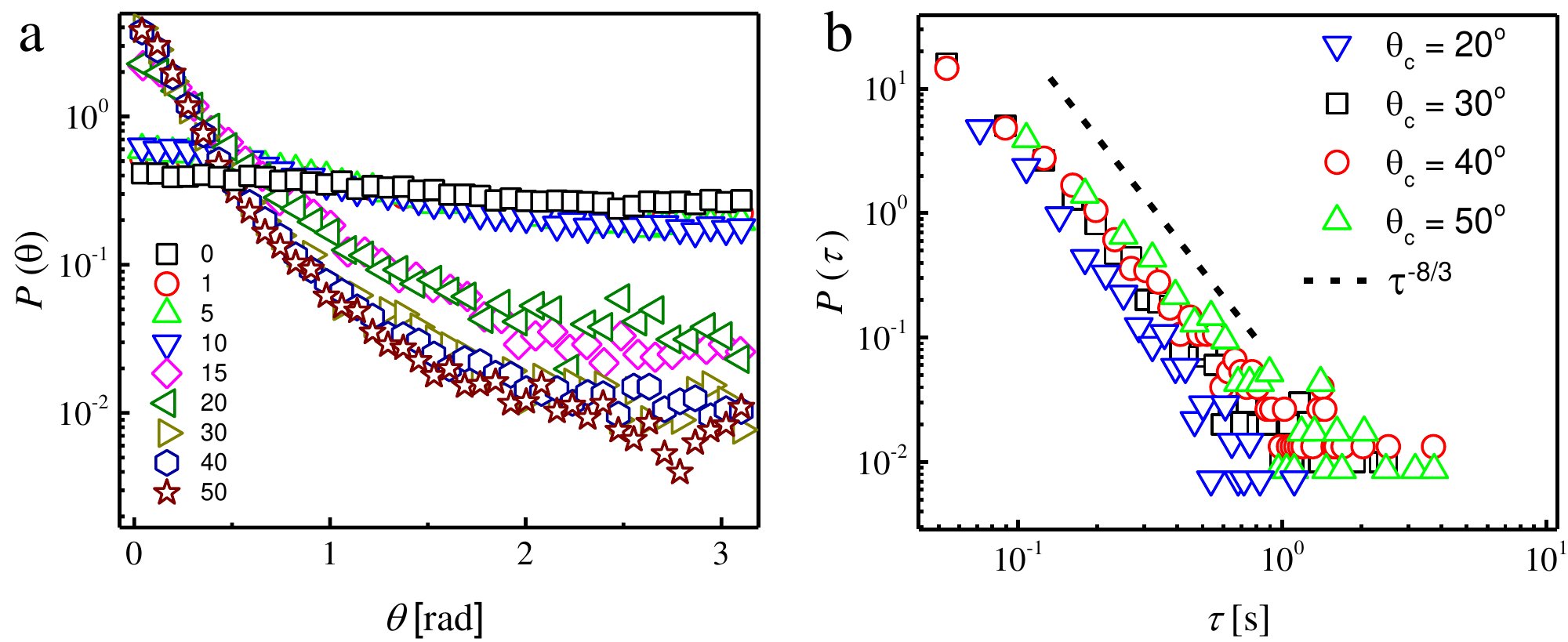}
    \caption{(a) Probability distribution of colloid's turning angles for bacterial concentrations. Without bacteria, the distribution is almost flat, indicating the random motion of colloids. Below 10 $c_{\rm{0}}$ bacterial concentration, the curve shows a deviation, with a slight increase in the probabilities of small angles. In the collective dynamics regime (after 15 $c_{\rm{0}}$), The probability of small turning angles further increases. With an increase in the concentration, the probability of sharp turns decreases. 15 and 20 $c_{\rm{0}}$ shows a high probability of sharp turns, indicating the levy walk-like statics. (b) Probability distribution of waiting time for four different selections of critical threshold angle ($\theta_{\rm{c}}$) for 15 $c_{\rm{0}}$. All the curves show similar scaling behavior, providing compiling evidence in favor of L\'evy walk.}
    \label{fig:s5}
\end{figure*} 
\begin{figure*}
    \centering
    \includegraphics[width = 10cm]{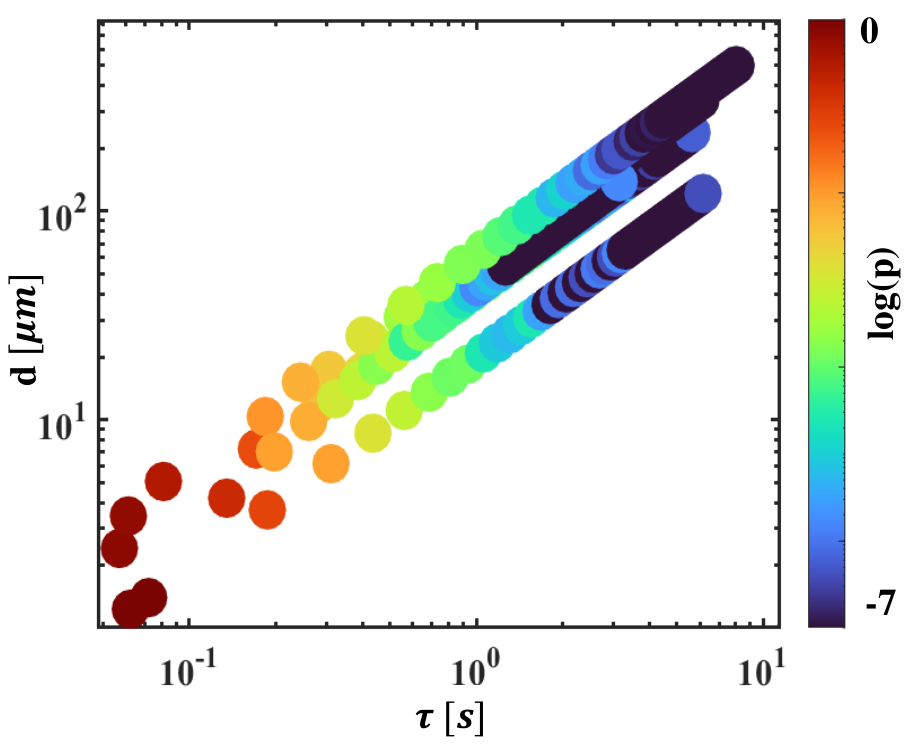}
    \caption{The joint probability of waiting time and step length, shows a linear scaling. This  indicates finite L\'evy velocity further supporting the presence of intermittent L\'evy walks in systems exhibiting collective dynamics.}
    \label{fig:s6}
\end{figure*}

\end{document}